\documentclass[pra,twocolumn,aps,showpacs,floatfix,superscriptaddress]{revtex4}

\usepackage{graphicx}
\usepackage{rotating}
\usepackage{amsmath}
\usepackage{bbm}
\usepackage{subfigure}
\usepackage{color}

\newcommand{\be}{\begin{equation}}
\newcommand{\ee}{\end{equation}}
\newcommand{\bea}{\begin{eqnarray}}
\newcommand{\eea}{\end{eqnarray}}

\begin{document}

\title{Ionization potentials and electron affinities from reduced density matrix functional theory
}

\author{E. N. Zarkadoula}
\affiliation{Theoretical and Physical Chemistry Institute, National Hellenic Research Foundation, Vass. Constantinou 48, GR-11635 Athens, Greece}
\affiliation{School of Physics and Astronomy, Queen Mary, University of London,
Mile End Road, London E1~4NS, United Kingdom}
\author{S. Sharma}
\affiliation{Max-Planck-Institut f\"ur Mikrostrukturphysik, Weinberg 2, D-06120 Halle, Germany}
\author{J. K. Dewhurst}
\affiliation{Max-Planck-Institut f\"ur Mikrostrukturphysik, Weinberg 2, D-06120 Halle, Germany}
\author{E. K. U. Gross}
\affiliation{Max-Planck-Institut f\"ur Mikrostrukturphysik, Weinberg 2, D-06120 Halle, Germany}
\author{N. N. Lathiotakis}
\affiliation{Theoretical and Physical Chemistry Institute, National Hellenic Research Foundation, Vass. Constantinou 48, GR-11635 Athens, Greece}
\begin{abstract}

In the recent work of S. Sharma \emph{et al.}, (arxiv.org: cond-matt/0912.1118), a single-electron spectrum
associated with the natural orbitals was defined as the derivative of the total energy
with respect to the occupation numbers at half filling for the orbital of interest.
This idea reproduces the bands of various periodic systems using the appropriate functional quite
accurately. In the present work
we apply this approximation to the calculation of the ionization potentials and electron affinities of
molecular systems using various functionals within the reduced density-matrix functional theory.
We demonstrate that this approximation is very successful in general and in particular for certain functionals
it performs better than the direct determination of the ionization potentials and electron affinities 
through the calculation of positive and negative ions respectively. 
The reason for this is identified to be the inaccuracy that arises from different handling of the open- and
closed-shell systems.

\end{abstract}

\date{\today}

\maketitle

\section{Introduction}
It is generally accepted today that the Fermi surfaces of metallic systems obtained 
with density functional theory (DFT), even at the level of local density approximation (LDA), are in good agreement with experiments. 
Unfortunately, this is not the case with the band gaps of insulators and semiconductors which are highly underestimated 
by most of the exchange-correlation (xc) functionals within DFT. 
Even with the exact xc functional of DFT, the Kohn-Sham (KS) gap is not expected to reproduce the experimental gap\cite{godby}.
This deviation from experiment is most dramatic for 
Mott-insulators, most of which are predicted by their KS spectrum to be metallic while they are experimentally known to be 
insulating in nature.

In this regard reduced density matrix functional theory (RDMFT) has shown great promise in improving on DFT 
results for a wide class of systems in that it not only improves the KS-band gaps for insulators in general, but also 
predicts the correct insulating nature for Mott insulators \cite{sharma08}.
Within RDMFT the total energy of a system of interacting electrons is expressed in terms of the one-body reduced density matrix 
(1-RDM), $\gamma({\bf r},{\bf r}')$. This energy functional is then minimized with respect to $\gamma$ under the $N$-representability 
conditions~\cite{coleman} which restrict the minimization to the domain of 1-RDMs that correspond to ensembles of $N$-electron 
wave functions. A major advantage of RDMFT comes from the fact that the {\it exact} kinetic energy is easily expressed as a functional
of the 1-RDM of the ground state. In addition, due to the departure from the
idempotent single-determinant solution, static electronic correlations are well described\cite{bb0}. The total ground-state energy
as a functional of $\gamma$ reads (atomic units are used throughout):
\begin{align} \label{etot} 
E[\gamma]=&-\frac{1}{2} \int\lim_{{\bf r}\rightarrow{\bf r}'}
\nabla_{\bf r}^2 \gamma({\bf r},{\bf r}')\,d^3{\bf r}'
+\int\rho({\bf r})\, V_{\rm ext}({\bf r})\,d^3{\bf r} \nonumber \\
&+\frac{1}{2}  \int 
\frac{\rho({\bf r})\,\rho({\bf r}')}
{|{\bf r}-{\bf r}'|}\,d^3{\bf r}\,d^3{\bf r}'+E_{\rm xc}[\gamma],
\end{align}
where $\rho({\bf r})=\gamma({\bf r},{\bf r})$. $V_{\rm ext}$ is a given external potential, and $E_{\rm xc}$ we call the xc 
energy functional. In practice, the xc functional is an unknown functional of the 1-RDM and needs to be approximated. 
A milestone in the development of approximate functionals of the 1-RDM is the M\"uller functional\cite{mueller,bb0}, 
which has the following form:
\begin{multline} \label{eq:mueller1} 
E_{\rm xc}[\gamma]=E_{\rm xc}[\{\phi_{j}\},\{n_{j}\}] = \\ -\frac{1}{2}\int  \int
 \frac{|\gamma^{1/2}({\bf r},{\bf r}')|^2}{|{\bf r}-{\bf r}'|} \, d^3{\bf r}\, d^3{\bf r}'
\end{multline}
where $1/2$ is an exponent in the operator sense.  Diagonalization of $\gamma$ produces a set of natural orbitals 
(the eigenvectors of $\gamma$), $\phi_{j}$, and occupation numbers (the eigenvalues of $\gamma$), $n_{j}$.
The  M\"uller functional is known to over correlate\cite{staro,herbert,gritsenko,frank,nekjellium},
 however, there exist several other approximations most of which are modifications of this 
functional and are known to improve results for finite systems \cite{GU,LHG2005,pnof1,pnof2,piris_os,gritsenko,helbig07,nekjellium,pernal_epot,LM2008,pade,pankratov,piris_os_n,AC3,Lathiotakis09,PNOF3,eich2010_prb,nhg2010_size,tolo2010,pernal2010,PNOF5}.

Several of these RDMFT functionals reproduce the discontinuity of the chemical potential at integer number of electrons 
which is a measure of the fundamental gap of the system\cite{helbig07,sharma08,helbig09,dfg10}. More precisely, it was 
demonstrated \cite{helbig07,helbig09,dfg10} that the complete 
removal of the self-interaction (SI) terms leads to discontinuities in the chemical potential that are in good agreement with the
fundamental gap for finite systems.  Unfortunately, this removal has no effect in the total energy
for infinitely extended natural orbitals in periodic systems since  their contribution vanishes in the limit of the size of 
system going to infinity. To overcome this problem, Sharma \emph{et al.} \cite{sharma08}
introduced the power functional\cite{sharma08,power_finite} that reproduces discontinuities without requiring the removal of SI terms. 
This functional has the form
\begin{multline} \label{eq:alpha} 
E_{\rm xc}[\gamma]=E_{\rm xc}[\{\phi_{j}\},\{n_{j}\}] = \\ -\frac{1}{2}\int \, \int 
 \frac{|\gamma^{\alpha}({\bf r},{\bf r}')|^2}{|{\bf r}-{\bf r}'|}\,d^3{\bf r}\, d^3{\bf r}'
\end{multline}
where $\alpha$ is an exponent in the operator sense.
The power functional was applied in the calculation of the fundamental gap of various systems\cite{sharma08,tolo2010} including
transition metal oxides \cite{sharma08}.
An optimal value of $\alpha$ between $0.6$ and $0.7$ was found to reproduce gaps of all systems in close agreement
with experiments. These gaps were obtained from the discontinuity of the chemical potential, $\mu(N)$,
at an integer total number of electrons $N$.
A problem of this method of predicting the gap is that the shape of $\mu(N)$ 
differs substantially from a step function leading to large error-bars in the prediction of the gap. 
A second problem is that one needs to calculate the total energy (and $\mu$) for several values of $N$ making the calculation 
time consuming. 
Finally, a third problem is that this method does not allow for the calculation of quantities other than the gap, 
for direct comparison with experiments, like for example the density of states for extended systems and 
ionization potentials (IPs) and electron affinities (EAs) for finite systems.

An advantage of DFT is that the KS eigenvalues can be used as an approximate 
single-electron spectrum of the system. Thus, quantities like the IP and EA can 
be easily estimated using the KS spectrum. A fundamental difference between RDMFT and DFT is the lack of a KS system within
RDMFT and the lack of eigenvalue  equation makes it difficult to obtain (even approximately) the IPs and the EAs.
One way to calculate IPs in RDMFT, is to use extended Koopman's theorem (EKT), as was proposed by 
Pernal and Cioslowski\cite{pernalip}. They demonstrated that the Lagrangian matrix in RDMFT is 
identical with the generalized Fock matrix entering EKT. Thus, IPs can be calculated by diagonalization of this
matrix. They used this idea in the calculation of IPs for small molecular systems
using the so called Buijse Baerends Corrected (BBC) \cite{gritsenko} and Goedecker Umrigar (GU)\cite{GU} functionals and showed that the error in the obtained IPs is of 
the order of 4-6\%. The same idea was employed in combination with yet another xc functional,
namely the PNOF1\cite{pnof1,pnof2} functional, for the calculation of the first IPs (FIPs) as well as higher IPs (HIPs)
of molecular systems yielding results of similar quality\cite{piris_calc}. However, the application of this method 
is restricted to finite systems, since, for solids, it would require the diagonalization of a large matrix in 
wave-vector space.  

Sharma \emph{et al.} in Ref.~[\onlinecite{sharmaarch}] introduced an alternative technique to obtain spectral information. 
We refer to this technique as the ``derivative'' (DER) method as it entails for each natural orbital, $k$, 
the associated energy, $\epsilon_k$, be obtained  as the 
derivative of the total energy with respect to the occupation number, $n_k$, at $n_k=1/2$ and with the rest of the occupation numbers 
set equal to their ground-state optimal values. This technique has been applied for 
the calculation of densities of states of transition-metal oxides (NiO, FeO, CoO and MnO)
and the results were found to be in excellent agreement with experiments \cite{sharmaarch} and other state-of-the-art many-body techniques like
dynamical mean-field theory and the $GW$ method. 

In the present work this technique is applied to finite systems. We discuss the validity of the approximations 
necessary for the accuracy of the method. We present results for the FIP as well as HIP, 
for atoms and molecules as well as the EAs of atoms, molecules, and radicals adopting several present-day 
functionals of the 1-RDM.  We compare the results with EKT, QCI(T), and experiment.

\section{theory}

\begin{table}[t]
\setlength{\tabcolsep}{0.1cm}
\begin{tabular}{lccc}
{\bf Funct.} &\bf  ${\bf \Delta_{\rm DEF}}$(\%) &\bf  ${\bf \Delta_{\rm DIF}}$(\%) &\bf  ${\bf \Delta_{\rm DER}}$(\%) \\ \hline\hline
M\"uller\cite{mueller} & 13.24 & 12.82 & 10.03  \\
GU\cite{GU}    &  5.90 & 10.19 &  9.19  \\
Power\cite{sharma08}    & 12.51 &  8.68 &  6.08  \\
AC3\cite{AC3}      &  5.98 &  9.33 &  6.33  \\
PNOF1\cite{pnof1,pnof2}     &  6.27 & 11.05 &  7.21  \\
BBC3\cite{gritsenko}     &  6.24 & 10.19 &  8.79  \\
ML\cite{pade}       &  6.15 &  9.22 &  4.17  \\
\hline
\end{tabular}
\caption{\label{tab:errors}
Average absolute errors $\Delta_{\rm DEF}$, $\Delta_{\rm DIF}$, and $\Delta_{\rm DER}$, in the calculation of IP, FIPs and HIPs, for a set of atoms and molecules with calculations performed using
various xc functionals in conjunction with DEF, DIF, DER methods respectively.}
\end{table}

By definition, the ionization potential and electron affinity are given by
\begin{equation}
   \begin{array}{lcl}
        {\rm IP} &=& E(N-1)-E(N)\\
        {\rm EA} &=& E(N)-E(N+1),\\
   \end{array}
   \label{eq:exIPEA}
\end{equation}
where $E(N)$ is the ground-state total energy of the charge-neutral system, and
$E(N-1)$, ($E(N+1)$) is the energy of the system with one electron removed (added). 
In the rest of the article we refer to this method of calculating the IP and 
EA as the definition method (DEF).
Due to Koopman's theorem, within the Hartree Fock (HF) theory, the IP in Eq.~(\ref{eq:exIPEA}) 
is well approximated by the eigenvalue of the highest occupied molecular orbital (HOMO). 
On the other hand, within DFT, the KS energy of the HOMO is exactly equal to the IP in Eq.~(\ref{eq:exIPEA})
for the exact xc functional. Likewise, the exact EA equals the orbital energy of the HOMO of the $N+1$
electron system calculated with the exact xc functional.

\begin{table*}[t]
\setlength{\tabcolsep}{0.2cm}
\begin{tabular}{llcccccccccc}
{\bf System } &      & {\bf M\"uller}  &{\bf  GU} &{\bf  Power} &{\bf  AC3} &{\bf PNOF1} &{\bf  BBC3} 
&{\bf  ML} & {\bf HF\footnotemark[1]}& \bf QCI(T) &{\bf  Expt. } \\ \hline\hline
He         & FIP & 25.361 & 25.252 & 25.579 & 24.953 & 24.254 & 25.307 & 24.626 & 24.871 & 24.327 & 24.59\footnotemark[2] \\
H$_2$      & FIP & 16.490 & 16.463 & 16.599 & 16.136 & 16.419 & 16.436 & 16.028 & 16.109 & 16.245 & 15.43\footnotemark[3] \\
LiH        & FIP & 8.191  & 8.381  & 8.490  & 8.136  & 8.307  & 8.408  & 8.027  & 8.109  & 7.782  & 7.78\footnotemark[4]  \\
H$_2$O     & FIP & 10.259 & 11.592 & 11.864 & 13.007 & 11.614 & 13.415 & 12.844 & 13.415 & 11.919 & 12.78\footnotemark[5] \\
           & HIP 1& 16.436 & 15.619 & 15.674 & 16.871 & 14.116 & 17.361 & 15.266 & 15.402 &        & 14.83\footnotemark[5] \\
           & HIP 2& 19.075 & 17.769 & 18.994 & 18.721 & 17.609 & 19.266 & 18.640 & 18.857 &        & 18.72\footnotemark[5] \\
HF         & FIP & 15.783 & 15.130 & 16.245 & 16.925 & 15.199 & 17.769 & 16.463 & 17.116 & 15.429 & 16.19\footnotemark[5] \\
           & HIP 1& 20.381 & 18.803 & 20.055 & 19.647 & 18.591 & 21.415 & 20.055 & 20.327 &        & 19.90\footnotemark[5] \\
CH$_4$     & FIP & 13.578 & 12.817 & 14.123 & 14.449 & 12.960 & 15.592 & 14.395 & 14.776 & 14.177 & 14.40\footnotemark[5] \\
           & HIP 1& 24.028 & 23.538 & 24.191 & 25.497 & 22.508 & 20.300 & 25.089 & 25.633 &        & 23.00\footnotemark[5] \\
CO$_2$     & FIP & 9.796  & 9.415  & 11.320 & 13.143 & 10.664  & 15.483 & 13.633 & 14.558 & 13.225 & 13.78\footnotemark[6] \\
           & HIP 1& 16.735 & 16.789 & 17.524 & 19.102 & 15.389 & 19.429 & 18.667 & 19.075 &        & 17.30\footnotemark[6] \\
NH$_3$     & FIP & 8.626  & 9.878  & 10.150 & 10.966 & 9.999  & 11.510 & 10.966 & 11.429 & 10.340 & 10.80\footnotemark[5] \\
           & HIP 1& 17.116 & 16.136 & 16.626 & 17.062 & 15.611 & 17.551 & 16.599 & 16.708 &        & 16.80\footnotemark[5] \\
Ne         & FIP & 20.898 & 20.272 & 21.443 & 22.585 & 20.319 & 23.347 & 21.824 & 22.640 & 20.871 & 21.60\footnotemark[7] \\
           & HIP 1& 48.028 & 47.484 & 48.980 & 52.600 & 46.629 & 52.899 & 51.130 & 52.219 &        & 48.47\footnotemark[7] \\
C$_2$H$_4$ & FIP & 6.748  & 8.218  & 8.299  & 9.361  & 9.674  & 9.796  & 9.714  & 10.177 & 10.422 & 10.68\footnotemark[8] \\
           & HIP 1& 11.674 & 14.068 & 12.708 & 14.232 & 13.642  & 14.340 & 13.660 & 13.660 &        & 12.80\footnotemark[8] \\
C$_2$H$_2$ & FIP & 11.048 & 10.721 & 10.966 & 10.939 & 9.821  & 11.538 & 11.402 & 10.966 & 11.184 & 11.49\footnotemark[5] \\
           & HIP 1& 21.198 & 19.701 & 19.783 & 19.130 & 16.751 & 20.136 & 20.028 & 18.340 &        & 16.70\footnotemark[5] \\
           & HIP 2& 21.143 & 22.041 & 20.653 & 21.633 & 18.302 & 20.653 & 20.626 & 20.789 &        & 18.70\footnotemark[5] \\ \hline
 $ \Delta_{\rm FIP}    $ (\%)& & 12.38 & 10.91 & 7.42 & 4.10 & 9.19 & 6.93 & 2.07 & 4.43 & 2.91 & 0.00\\
 $ \Delta_{\rm HIP}$ (\%)& & 7.45 & 7.29 & 4.61 & 8.78 & 5.04 & 10.83 & 6.48 & 6.43 &   &  0.00\\
 $ \Delta      $ (\%)& & 10.03 &  9.19 &  6.08 &  6.33 & 7.21 & 8.79  &  4.17 & 4.45 &   &  0.00 \\
\hline
 $ \Delta^{\rm (EKT)}_{\rm FIP}$ (\%)& & 12.22 & 2.47 & 6.26 & 2.82 & 2.99 & 5.16 & 5.11 &  &  & \\
 $ \Delta^{\rm (EKT)}_{\rm HIP}$ (\%)& & 10.56 & 5.69 & 11.80 & 6.45 & 5.15 & 4.21 & 12.02 & & & \\
 $ \Delta^{\rm (EKT)}      $ (\%)& & 11.43 &  4.00 &  8.90 &  4.55 & 4.02 & 4.71  &  8.40 &  & & \\
\hline
\end{tabular}
\footnotetext[1]{with Koopman's theorem (${\rm IP}=-E_{\rm HOMO}$)}
\footnotetext[2]{Ref.~[\onlinecite{g2set2}]}
\footnotetext[3]{Ref.~[\onlinecite{eyler}]}
\footnotetext[4]{Ref.~[\onlinecite{IW1975}]}
\footnotetext[5]{Ref.~[\onlinecite{piris_calc}]}
\footnotetext[6]{Ref.~[\onlinecite{CO2spectr}]}
\footnotetext[7]{Ref.~[\onlinecite{cederbaum}]}
\footnotetext[8]{Ref.~[\onlinecite{bieri}]}
\caption{\label{tab:ip}
Ionization potentials (FIPs and HIPs), in eV, for various molecules calculated with different functionals using DER method.  
These results are compared with Hartree-Fock, QCI(T), and the experimental data. 
QCI(T) values were calculated with Gaussian 09 program\cite{g09} using the same basis set
through Eq.~(\ref{eq:exIPEA}). In the bottom row are included the percentage absolute average errors $\Delta_{\rm FIP}$, 
$\Delta_{\rm HIP}$, and $\Delta$ in the calculation of the FIPs, HIPs and all IPs respectively. 
For comparison, the errors $\Delta^{\rm (EKT)}_{\rm FIP}$, 
$\Delta^{\rm (EKT)}_{\rm HIP}$, and $\Delta^{\rm (EKT)}$ using the EKT are also included.}
\end{table*}

Within RDMFT, there is no effective single particle KS system reproducing
the non-idempotent 1-RDM of the interacting system and quantities like IP and EA can not 
be obtained from an eigenvalue equation. However, approximate but meaningful  
single particle energies associated with the natural orbitals can be defined as in Ref. [\onlinecite{sharmaarch}]:
\begin{equation}\label{eq:dif}
   \epsilon_k = \left. E(\{n_j\})\right|_{n_k=1} - \left. E(\{n_j\})\right|_{n_k=0}.
\end{equation}
The two energies on the right hand side are the energies of the system with all natural orbitals and occupation numbers having
the optimal ground-state values except for the natural orbital of interest, $k$, for which occupation numbers are set to either $n_k=1$ or $n_k=0$. 
In this way, these energies are approximate electron addition and/or removal energies for the natural orbital $k$. We refer to 
this method for calculating IPs and EAs as the ``energy-difference'' method (DIF).

It has been shown that, for extended systems \cite{sharmaarch}, 
the total energy is almost linear if a particular occupation, $n_k$, is 
varied between zero and 1. If it was exactly linear then the energy difference in Eq.~(\ref{eq:dif}) would be given by the
tangent of  $E(\{n_j\})$. In absence of this linearity a good choice is to use the Slater trick and approximate $\epsilon_k$ by 
\begin{equation}
\label{eq:spectrum}
\epsilon_k = \left. \frac{\partial E[\{\phi_j\},\{n_j\}]}{\partial n_k} \right|_{n_k=1/2},
\end{equation}
where the derivative is calculated at the ground-state natural orbitals $\{\phi_j\}$ and 
occupations $\{n_j\}$, except for $k$  which  is set to  $n_k=1/2$. This is a good approximation because if one expands 
$E(\{n_j\})$ at $n_j=1/2$ the term in Eq.~(\ref{eq:spectrum}) is the leading order term with the second order 
term being identically equal to zero.
At first sight Eq.~(\ref{eq:spectrum}) looks similar to the Janak's theorem\cite{janak}, 
which gives the eigen-energies of the Kohn-Sham system within DFT. However, it is important to note that
within RDMFT lack of single particle eigenvalue equations does not permit the direct use of Janak's theorem--
Janak's theorem would lead to all orbital energies, for fractionally occupied states, to be degenerate with value equal 
to the chemical potential.

\section{methodology\label{sec:method}}

\begin{figure*}[t]
\begin{center}
\begin{tabular}{cc}
\includegraphics[width=0.47\textwidth,clip]{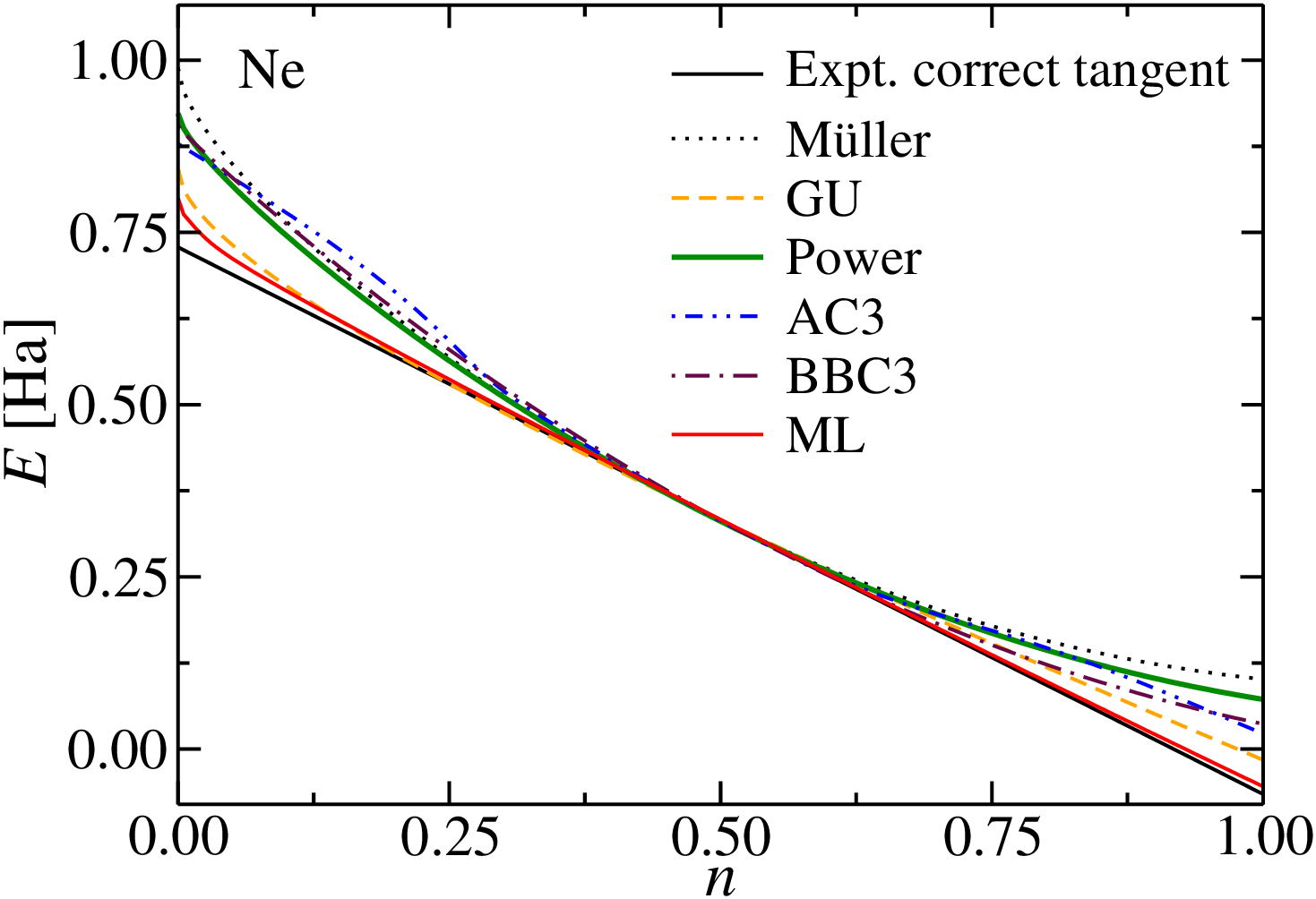} & \includegraphics[width=0.47\textwidth,clip]{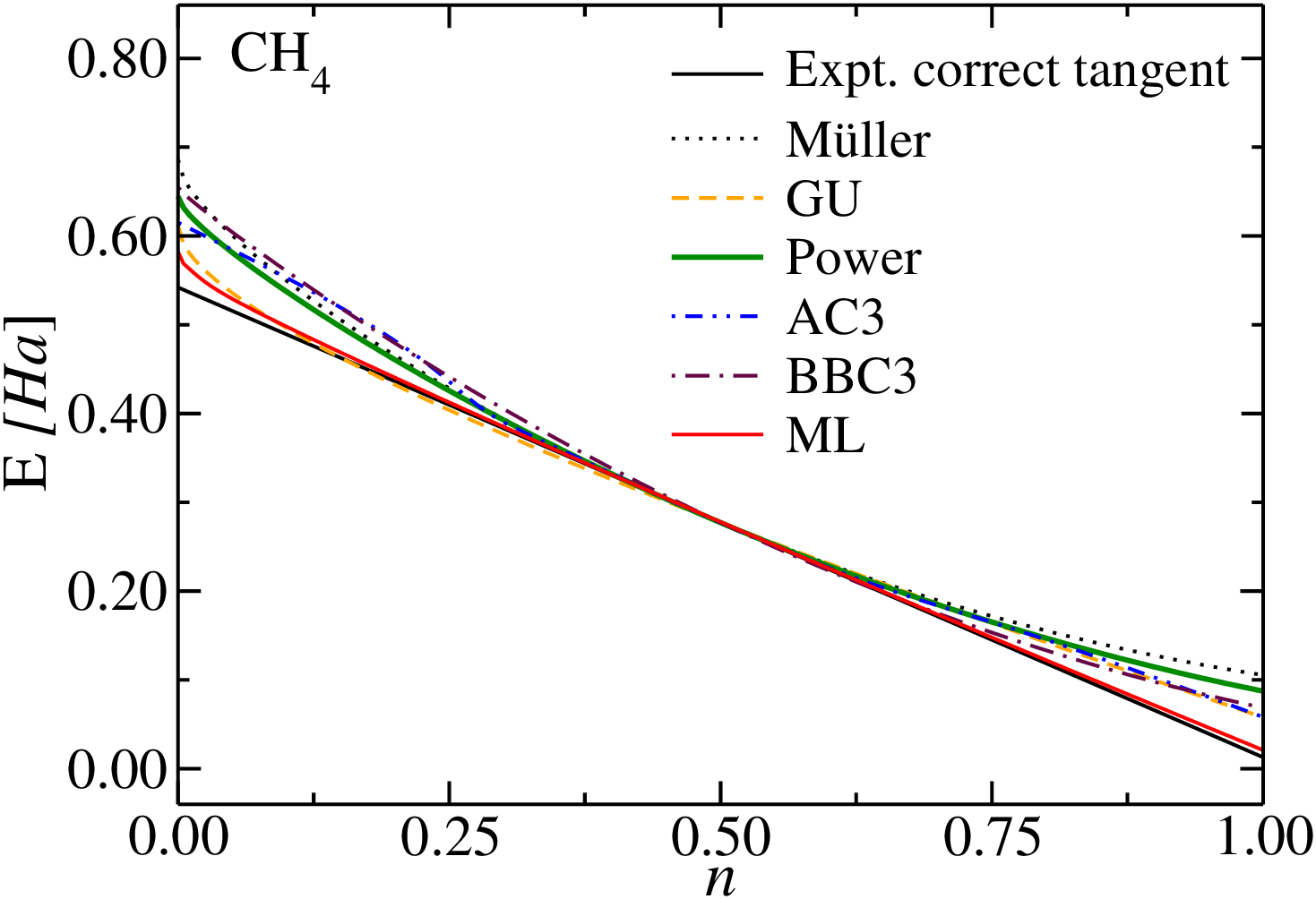}\\[0.3cm]
\includegraphics[width=0.47\textwidth,clip]{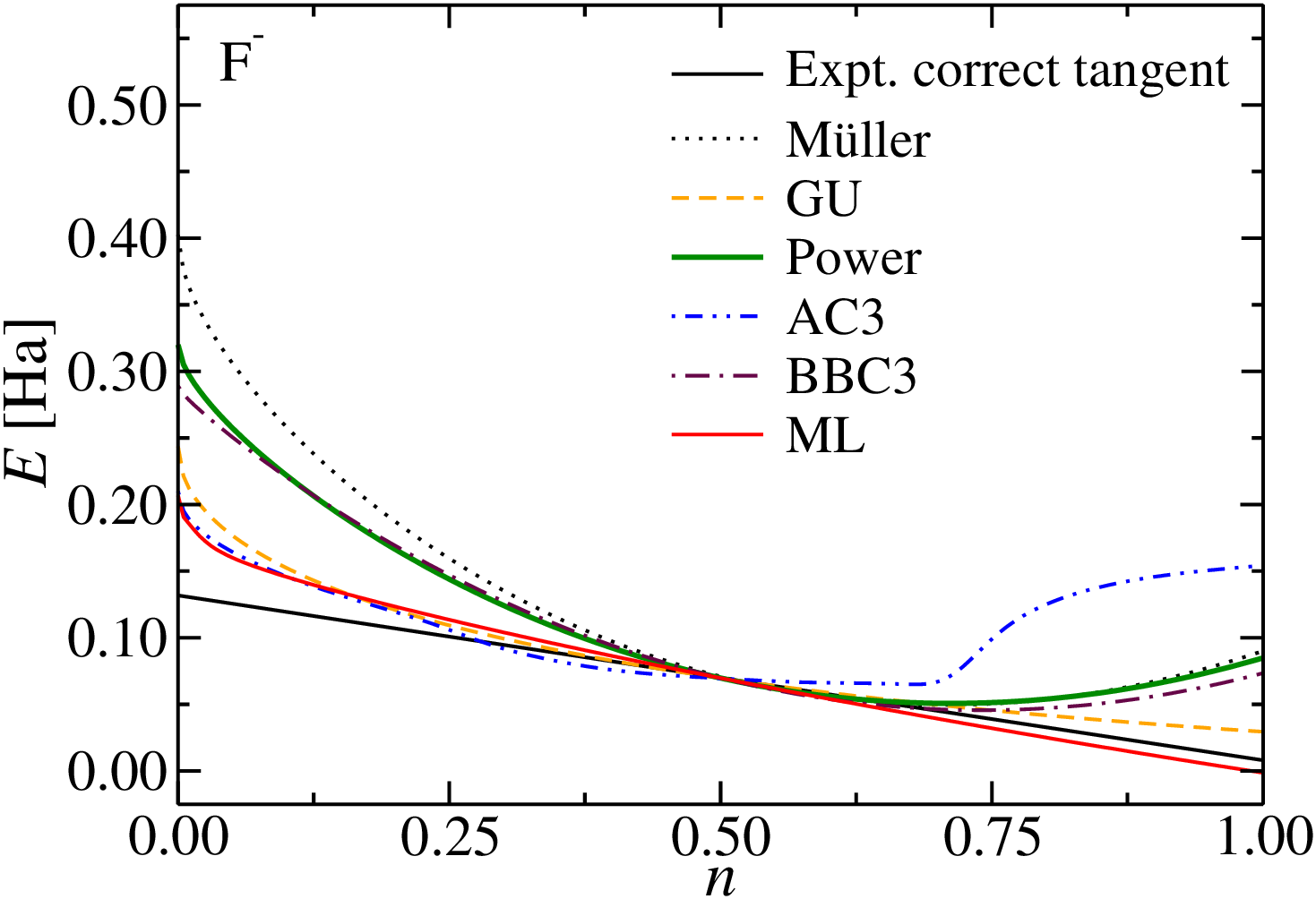} & \includegraphics[width=0.47\textwidth,clip]{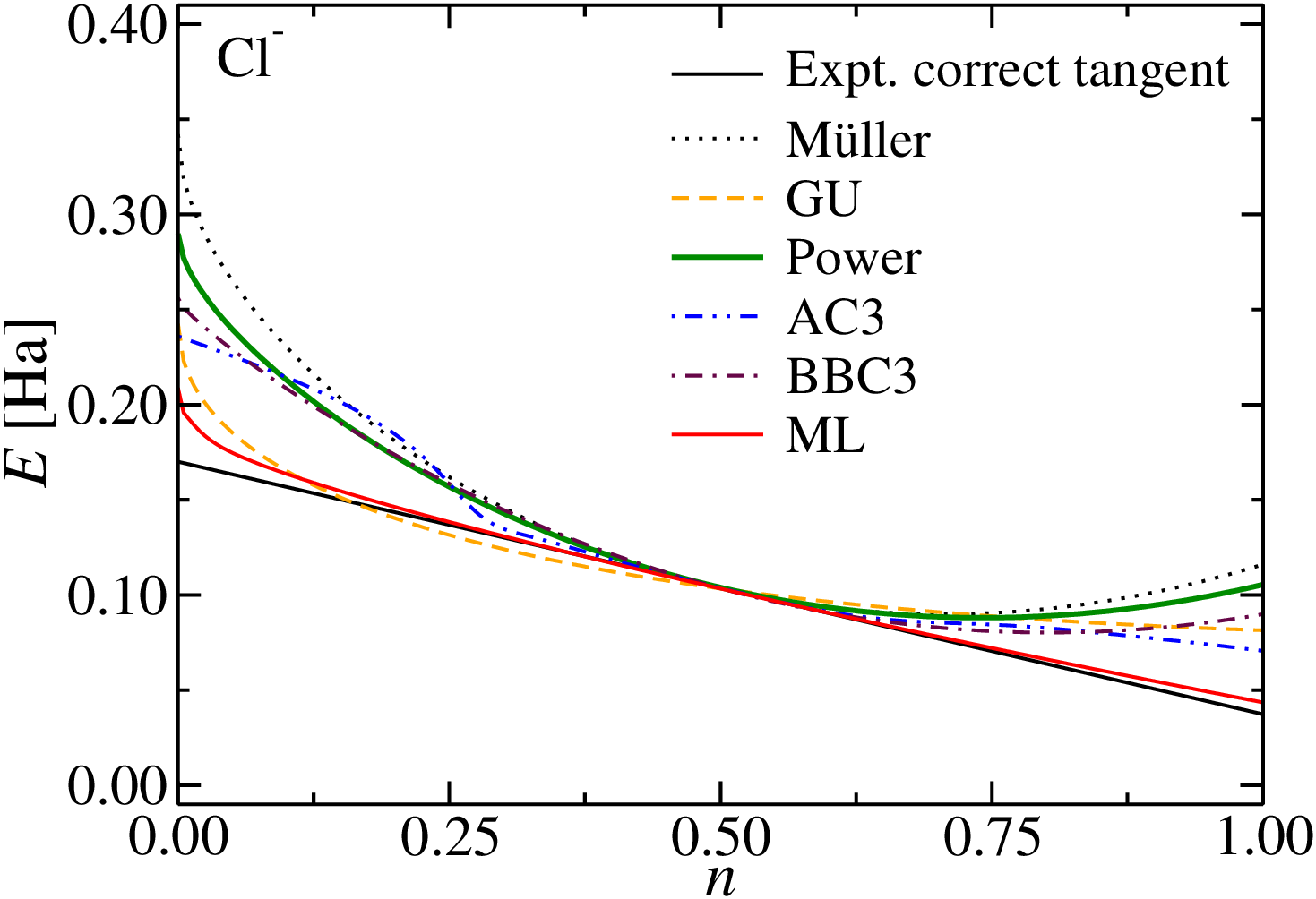} \\
\end{tabular}
\end{center}
\caption{\label{fig:E_nIP}(Color online) The total energy, $E$, as a function of the occupation number, $n$, of the HOMO, for Ne atom
and CH$_{\rm 4}$ (top) and the negative ions F$^-$ and Cl$^-$ (bottom). 
The line  giving the correct experimental values for IP/EA is also shown.
Curves are shifted to coincide at $n=1/2$ in order to compare the tangents with the straight line reproducing the 
experimental results. The values at the two ends are used in DIF for the 
calculation of IP for Ne,  CH$_{\rm 4}$ and the affinity of F, Cl. The derivatives at $n=1/2$ are used for the calculation of 
the same quantities with the DER method.}
\end{figure*}

We calculate the IPs and EAs of a set of atoms and molecules
using DEF, DIF, and DER methods [i.e., using the Eqs.~(\ref{eq:exIPEA}), (\ref{eq:dif}) and (\ref{eq:spectrum})]. 
For comparison, results are also calculated using the EKT method. Our implementation is included in 
a computer code for finite systems\cite{code} which minimizes 1-RDM functionals with respect
to occupation numbers and natural orbitals and is based on the expansion of the orbitals in Gaussian basis sets.
The one- and two-electron integrals are calculated by use of the GAMESS program\cite{gamess}. Addition or removal of an
electron requires the extension of the theory to open shells. In the present work, like in 
Refs.~[\onlinecite{helbig07,helbig09,dfg10,fracspin}] we use the simple extension proposed in Ref.~[\onlinecite{LHG2005}]. In other words,
we assume that orbitals are spin independent while occupation numbers are spin dependent. 

For the calculation of 
IPs we adopt the cc-pVDZ basis set\cite{ccpvdz} for all elements. EAs are calculated as the IPs of negative ions, i.e.,
of $N+1$ electrons. In other words, for both IPs and EAs, the orbital energy of the HOMO (for either the neutral or
ionic system) is calculated. Since the HOMOs of the negative ions are relatively diffuse states, the aug-cc-pVDZ
basis set is used~\cite{ccpvdz}. We should mention that a lot of neutral systems do not bind an extra electron.
In that case, EA is equal to zero, i.e. the extra electron is completely delocalized. However, for small positive 
EA, the state of the extra electron can be delocalized and impossible to describe with localized basis sets. To ensure 
fair comparison with experiments a set of atoms, molecules and radicals which are known experimentally to have relatively
large and positive EA is used.

Calculation of IPs and EAs with DEF method requires  the difference of two energies, one for a 
closed-shell and another for an open shell-system. The broken spin symmetry in open-shell systems leads to twice as many variational
parameters as there are in a closed shell system. This extra variational freedom over correlates the 
open-shell systems leading to systematic errors in IPs and EAs. Given the exact xc functional the DEF method would be exact, however, 
for an approximate functional the DEF method would show these systematic errors in IPs and EAs.
DIF and DER methods on the other hand do not suffer from this error since only one minimization, for the charge neutral system, is performed.
In addition, it should not come as a surprise if 
the DER method performs better than DEF and DIF in many cases
as it suffers less from possible inaccuracies introduced by the functional and its non unique 
extension to the case of open shells, mainly because only the $1/2$ electron is present in the open shell.
One could also consider DER method in conjunction with orbital relaxation (at fixed $n_j=1/2$).
However, this procedure requires full orbital minimization for each $j$ making it computationally very demanding, while the
aim of the present work is to define 
a computationally inexpensive single electron spectrum in terms of the optimal 1RDM of the charge neutral system.

\begin{table}[h]
\setlength{\tabcolsep}{0.2cm}
\begin{tabular}{lccc}
{\bf Functional } & {\bf $\bf \Delta_{\rm DEF}$(\%)} &{\bf $\bf \Delta_{\rm DIF}$(\%)} & {\bf $\bf \Delta_{\rm DER}$(\%)} \\
\hline\hline
M\"uller & 69.85 & 88.55  &  63.24 \\
GU       & 44.91 & 65.96  &  70.81 \\
Power    & 64.69 & 63.57  &  48.73 \\
AC3      & 42.40 & 31.17  &  21.39 \\
PNOF1    & 39.68 & 61.44  &  28.63 \\
BBC3     & 30.00 & 55.33  &  52.00 \\
ML       & 39.32 & 47.81  &  21.53 \\
HF       & 55.00 &        &        \\	
CI/QCI(T)& 17.28 &        &        \\
\hline
\end{tabular}

\caption{\label{tab:eaerr}
Average absolute errors $\Delta_{\rm DEF}$, $\Delta_{\rm DIF}$, and $\Delta_{\rm DER}$, in the calculation of EAs for a set of atoms, molecules and radicals calculated with various xc functionals using DEF, DIF, and DER respectively.
}
\end{table}

Another point to be considered is that the application of the DER method requires the total energy functional to be
continuous at $n_k=1/2$.  However, there are functionals that introduce a discontinuity at  $n_k=1/2$ 
to distinguish between strongly and weakly occupied orbitals\cite{piris_calc}. In all cases studied here, we do not find
optimal occupation numbers equal to 1/2. Thus, the step function can be safely shifted slightly away from $n_k=1/2$   
without affecting the results. However, this procedure cannot be used in cases with optimal occupations equal to 1/2, like
H$_2$ at the dissociation limit or when they vary continuously from 1 to zero.

\begin{table*}[t]
\setlength{\tabcolsep}{0.1cm}
\begin{tabular}{lccccccccc}
{\bf System } &  {\bf M\"uller}  &{\bf  GU} &{\bf  Power} &{\bf  AC3} & {\bf  PNOF1} &{\bf  BBC3} &{\bf  ML} & \bf QCI(T) &{\bf  Expt. } \\ \hline\hline
LiH   & 0     & 0.192 & 0     & 0.219 & 0.317  & 0.106 & 0.264 & 0.317 & 0.34\footnotemark[1] \\
OH    & 3.157 & 6.735 & 3.456 & 2.626 & 1.784  & 3.157 & 2.441 & 1.645 & 1.83\footnotemark[2] \\
F     & 5.651 & 3.297 & 5.040 & 3.196 & 3.552  & 5.539 & 4.363 & 3.241 & 3.34\footnotemark[2] \\
Li    & 0     & 0.403 & 0.139 & 0.275 & 0.400  & 0     & 0.212 & 0.601 & 0.62\footnotemark[2] \\
Cl    & 3.687 & 2.190 & 3.632 & 3.678 & 2.660  & 4.186 & 3.623 & 3.430 & 3.61\footnotemark[2] \\
CN    & 1.807 & 2.328 & 2.332 & 4.264 & 2.470  & 4.873 & 4.585 & 3.627 & 3.77\footnotemark[2] \\
C$_2$ & 1.044 & 1.419 & 1.298 & 3.386 & 4.416  & 4.560 & 3.722 & 3.055 & 3.54\footnotemark[2] \\
BO    & 1.243 & 1.326 & 1.982 & 3.252 & 2.076  & 3.222 & 3.252 & 2.359 & 2.83\footnotemark[3] \\
SH    & 1.386 & 0.706 & 1.908 & 1.968 & 1.352  & 2.454 & 2.168 & 2.119 & 2.32\footnotemark[2] \\
PH    & 0.304 & 0     & 1.200 & 1.252 & 0.186  & 2.192 & 1.145 & 2.023 & 1.00\footnotemark[3] \\ \hline
$\bf \Delta$\bf (\%) & 63.24 & 70.81 &  48.73  & 21.39 & 28.63 & 52.00 & 21.53 & 17.28 \\
\hline
\end{tabular}
\footnotetext[1]{Ref.~[\onlinecite{LiHspectr}]}
\footnotetext[2]{Ref.~[\onlinecite{ea_calc}]}
\footnotetext[3]{Ref.~[\onlinecite{piris_os}]}
\caption{\label{tab:ea}
Electron Affinities for various atoms, molecules and radicals calculated as 
the IP of the system of $N+1$ electrons with different functionals using the DER method. 
For systems where $N+1$ electrons energy is found
higher than the $N$ electron energy, zero affinity is assumed. The results are compared
with QCI(T) and experiments. QCI(T) values were calculated with
Gaussian 09 program\cite{g09}.}
\end{table*}

\section{results}
Our results for the average absolute errors in the calculation of IPs with the three methods are included in Table~\ref{tab:errors}. 
Average errors in the results obtained with EKT method are also included in the table.
The actual values for IPs obtained using the DER method are compiled in Table~\ref{tab:ip}.
(Full results for all methods as well as EKT can be found in the supplementary material\cite{supplem}.) It is clear from 
Table~\ref{tab:errors}  that all functionals in combination with the DER method give reasonable results for IPs with errors 
ranging from 4 to 13\%. 
ML, AC3 and power functionals perform slightly better by giving an average error of only 4-6\%.
For the  systems considered here, the ML functional with the DER method is the most accurate for the FIPs 
(with an error of only 2\%). 
It is important to note that the errors from the DER and EKT method are of the same order
(using the same functionals and basis set), while there is less computational effort involved in DER method.
The comparison of the DIF and DER methods  allows us to assess the validity of the linear approximation--the linearity of
the total energy with respect to variation of one occupation number while keeping the rest of the occupation numbers as well
as natural orbitals frozen (this was demonstrated for solids in Ref.~[\onlinecite{sharmaarch}]). As we see 
in Table~\ref{tab:ip}, the DER method gives good results also for finite systems and 
for the best performing functionals the average difference between the DIF and DER methods' results is in the range of 2-7\%. 
This percentage may be regarded as the magnitude of non-linearity of the total energy with respect to variation 
of a single occupation number.

As mentioned in Sec. \ref{sec:method}, the DEF method suffers from the over correlation error 
due to the approximate nature of the xc functional and the difference in variational freedom
between closed and open-shell systems. Since the DER method is less prone to this error, it is not a surprise 
that for the power and ML functionals the DER method improves 
the results over the DEF method. This indicates that the dependence of the total energy on a particular occupation number 
deviates from linearity but this deviation works in favor of the DER method by further improving the results. 
In order to understand this, in Fig.~\ref{fig:E_nIP}(top), we show the tangents at 1/2 of the total 
energy as a function of the occupation number of the HOMO, while keeping the rest frozen. 
The plots are made for various functionals.
It is clear from  Fig.~\ref{fig:E_nIP} that although the total energy functional itself deviates from 
linearity the tangent at 1/2 is very close to the one that reproduces the experimental results.  
 One reason for this improvement over the DIF method might be that the extension of the theory to open shells in the case the DER method is minimal-- 
 since only half of an electron is unpaired-- and DER method reduces the error introduced by the extension 
 of functionals to open shells. 

The average percentage errors and the values of HIPs for various atoms and molecules are also presented in Table~\ref{tab:errors}
and Table~\ref{tab:ip} respectively.
The average error in HIPs obtained using the M\"uller, GU, power and PNOF1 functionals, is substantially lower (4-7\%) than for the FIPs. 
The rest of the functionals are less accurate for HIPs with average absolute errors slightly higher than those for 
the FIPs (5-10\%).  

The average absolute errors in the calculation of EAs with the DEF, DIF, and DER methods are shown in Table~\ref{tab:eaerr}.
The actual values for EAs obtained using DER method are included in Table~\ref{tab:ea}. As 
already mentioned, EAs are more difficult quantities to calculate-- the errors are introduced by  
describing negative ions with localized basis which are usually optimized for the description of the ground states
of neutral systems. In addition, being a small quantity, EAs are more prone to errors in the differences
of total energies corresponding to two different shell structures (for example the difference in energy between a doublet and a 
singlet state). Thus, it is not surprising that the errors in Tables~\ref{tab:eaerr} and ~\ref{tab:ea} are substantially higher
than for the IPs. However, state-of-the art quantum chemical methods like QCI(T) also exhibit large errors within the adopted basis
set.  
Under these considerations the AC3, PNOF1 and ML functionals perform surprisingly well for EAs with average errors 
of 21.4\%, 28.6\%, and 21.5\% respectively. These errors are close to that of QCI(T) (17.3\%). 
It is interesting to note that (see Table~\ref{tab:ea}) there are only a few cases (5 of 60)
for which M\"uller, GU, BBC3 and power functionals give a zero EA [$E(N+1)>E(N)$], 
i.e., the system is not predicted by the corresponding functional to bind an extra electron. 
For the best performers like the AC3, PNOF1, and ML functionals, no such case exists.  

In order to compare the DIF and DER methods for the determination of EAs, 
Fig.~\ref{fig:E_nIP}(bottom) shows the tangents at 1/2 to the dependence of the total
energy on the occupation of the HOMO of the negative ions F$^-$ and Cl$^-$. The exact tangent that reproduces the
experimental EAs is also shown in the figure. Again, as in the case of IPs, the DER method not only
looks like a reasonable approximation but it also improves on the results of the DEF method (for the functionals
considered and in all cases studied in the present work).
In particular, for the case of Cl$^-$ [see Fig.~\ref{fig:E_nIP}(bottom)] the tangents at
1/2 are in very good agreement with the exact tangent that reproduces experimental EA, 
although the dependence of the total energy on the HOMO occupation number deviates significantly from linearity. 

A striking example of the pathological behavior mentioned in Sec.~\ref{sec:method} is the 
negative ion F$^-$.
This system is found experimentally to be energetically lower than the neutral F atom by 3.34~eV (see Table ~\ref{tab:ea}). 
All functionals underestimate this energy substantially (see supplementary material \cite{supplem}), as
a result of the enhanced variational freedom of the open-shell, neutral F atom compared to the closed-shell F$^-$.
In two extreme cases (ML and AC3 functionals) F$^-$ is found energetically above the neutral F atom. 

\section{summary}

In summary we examined the performance of the derivative-method proposed in Ref.~[\onlinecite{sharmaarch}]
for calculation of the IPs and EAs of finite systems. The accuracy of IPs and EAs calculated using the derivative-method 
are compared to the IPs and EAs calculated using the definition of these quantities [which involves two total energy 
minimizations for the system and the positive (for IP) or negative ion (for EA)]. In order to have a complete analysis we 
have also considered an intermediate method (difference method), in which the IPs and EAs are determined by the difference 
of the total energies with fixing one occupation number to 1 and/or zero. All these results are further compared to the 
state-of-the-art CI method as well as experiments. 

We find that, for IPs both the difference and derivative methods are good approximations to the definition of this 
quantity. Furthermore, it was found that the derivative method results, obtained using M\"uller, power, and ML functionals,
are better than the values obtained using the definition method itself (with errors of the order of 4-8\% only). 
Among these functionals the ML functional in conjunction with derivative method is most accurate with errors of only up to 2\%. 
For the EAs the errors are significantly larger, with the ML functional in conjunction with the derivative method being the most accurate 
(with an error of 21\%). The errors in EAs were found to be comparable to the errors in the CI results.
 
From the present study we conclude that the results of the derivative method for IPs and EAs are in good agreement with experiments
and this method is a promising technique to obtain the single-electron spectrum for systems where state-of-the-art quantum chemical
methods can not be applied and DFT results deviate significantly from experiment.


\end{document}